\documentclass{ptptex}

\def\ttau{{\tilde \tau}}
\def\half{\frac{1}{2}}
\def\bbz{{\mathbb Z}}
\def\cB{{\mathcal B}}
\def\cR{{\mathcal R}}

\def\cF{{\mathcal F}}
\def\ch{{\rm ch}}
\def\Irr{{\rm Irr}}
\def\Vir{{\rm Vir}}

\def\bt{{\mathbf T}}
\def\su2{\widehat{su}(2)_1}
\newcommand{\overlap}[2]{#1 q^{H_c} #2}
\newcommand{\ket}[1]{|#1\rangle}
\newcommand{\dket}[1]{|#1\rangle\!\rangle}
\newcommand{\bra}[1]{\langle #1|}
\newcommand{\dbra}[1]{\langle\!\langle #1 |}
\newcommand{\fusion}[4][\mathcal {N}]{{#1}_{#2#3}{}^{#4}}

\notypesetlogo                       
\preprintnumber[3cm]{
}

\title{
Conformal Boundary States in $\su2/G$%
}

\author{
Atsushi 
\textsc{Yamaguchi}\footnote{E-mail: ayamagu@post.kek.jp}
}

\inst{
Theory Division, High Energy Accelerator Research Organization (KEK),\\
Tsukuba, 305-0801, Japan
}

\recdate{March 1, 2004}

\abst{
We construct boundary states 
in a particular $c=1$ conformal field theory, the $\su2/G$ orbifold with
 $G$ a binary finite subgroup of $SU(2)$.
These states preserve the conformal symmetry, at least, but break 
rational symmetries of the $\su2/G$ orbifold in general.
}

\begin{document}
\maketitle
\section{Introduction and summary}
In the study of conformal field theories with boundaries, it is
important to construct 
boundary states generally breaking
extended symmetries but preserving 
the (reduced) conformal symmetry.
This problem has been clarified in $c=1$ irrational theories, in
particular, the circle
theory\cite{C_n}\tocite{C_n7} and its $\bbz_2$ orbifold
\cite{C_n7,Oshikawa:1996dj}.
In another $c=1$ CFT realized by the $\su2/G$
orbifold\cite{Ginsparg:1987eb}, 
the authors of Ref.~\citen{Cappelli:2002wq} constructed
the boundary states preserving the so-called automorphism-type rational symmetries.
After that work, a set of boundary states, 
which generally break such rational symmetries but preserve the
conformal symmetry, were constructed in Ref.~\citen{Ishikawa:2003xh} in the case that
$G$ is the order-$8$
dihedral group.
Our purpose here is to generalize this result to all 
binary finite subgroups of $SU(2)$. 

In this note, we construct a set of conformal 
boundary states (in the $\su2/G$
orbifold) characterized by $G$-orbits of the 
$SU(2)$ Lie group constituting 
a parameter space of prototypical
conformal boundary states in the $\su2$ WZW model, the parent theory of
the orbifold.
With respect to stabilizers in $G$, which fix these $G$-orbits,
our boundary states are classified into two types, referred to as 
a bulk type and a
fractional type\cite{Oshikawa:1996dj,Diaconescu:1999dt}. 
The bulk type boundary states exist on the
 $G$-orbits with the trivial stabilizers including only the
 identity and the center.
The fractional type boundary states exist on 
the $G$-orbits with non-trivial stabilizers and also correspond to
irreducible representations of these stabilizers. 
An important property is that a bulk type boundary
state extended to an 
arbitrary $G$-orbit with a nontrivial stabilizer is no longer
fundamental but splits into a linear combination 
of the fractional type boundary
states existing on this particular $G$-orbit. 
Among the boundary states mentioned above, some 
preserve rational symmetries on special $G$-orbits (although most break these symmetries).   
In particular, the Cardy
states\cite{Cardy:ir}, which preserve the symmetry of the full
 chiral orbifold algebra $\su2/G$, correspond to the
 conjugacy classes of $G\subset SU(2)$. 

This note is organized as follows. In the next section, we briefly review some aspects of boundary states in CFTs.
There we explain Cardy's conditions\cite{Cardy:ir} and the notion of fundamental
boundary states.
In \S\ref{bssu2}, we construct boundary states in the
 $\su2/G$ orbifold. First, we introduce boundary states in the $\su2$ WZW model.
Modding out these boundary states using the symmetry $G$, 
we next 
construct bulk type boundary  states existing on the $G$-orbits 
with the
trivial stabilizers. 
Then, we explain
that bulk type boundary states extended to $G$-orbits with
non-trivial stabilizers are
not fundamental.  
Non-fundamental boundary states are generally decomposed into linear
combinations of fundamental boundary states.\footnote{Non-fundamental
states are also relevant in statistical physics context (see, {\it
e.g.}, Ref.~\citen{aff} and references therein).}
We attempt to construct such additional fundamental boundary states,  which we refer to
as the ``fractional type'', with the following ansatz.
We start from consistent boundary states, for example, the Cardy states.
Next, we decompose them into
linear combinations of the Virasoro Ishibashi states\cite{Ishibashi:1988kg}.
Then, we deform the coefficients of these linear combinations to 
obtain expressions of more generic boundary states.
This deformation is carried out so that the resulting boundary states satisfy a set of strong constraints known as Cardy's conditions.
In this way, we can obtain other boundary states which are to be
identified with fractional type boundary states.
At the end of \S \ref{bssu2}, we mention which boundary states in
this note preserve rational symmetries in the $\su2/G$ orbifold.
In the Appendix, we give some technical details.

\section{Boundary states and Cardy's conditions}
Here, we give a short review of boundary states
 in a generic CFT with central charge $c$. 

Let us consider boundary states $\ket{B}$ that belong to a set denoted
by $\cB$ ({\it i.e.} $\ket{B}\in \cB$) 
and 
satisfy
the conformally invariant ``gluing'' condition, 
\begin{equation}
(L_n-\widetilde{L}_{-n})\ket{B}=0\ .   
\end{equation}
These boundary states are decomposed 
into linear combinations of the Ishibashi states
\cite{Ishibashi:1988kg} $\dket{j}^\Vir$, with $j$ being the
spin-less
irreducible representations of the left and right Virasoro algebras.
We can normalize 
the Virasoro Ishibashi states $\dket{j}^\Vir$ to 
satisfy the relations
\[
\overlap{{}^\Vir\dbra{j}}{\dket{j'}^\Vir}=\delta_{jj'}\chi_j(\tau)\ ,   
\]
where
 $q=e^{2\pi i \tau}$, $H=1/2(L_0+\widetilde{L}_0-c/12)$, and 
$\chi_j$ denotes the (left or right) Virasoro characters of
$j$.
Now, let us decompose the boundary states $\ket{B}\in \cB$ as follows:
\[
 \ket{B}=\sum_{j}C_B^j\dket{j}^\Vir\ .  
\]
The coefficients $C_B^j$ must satisfy 
a set of consistency conditions referred to as the sewing constraints (see
Refs.~\citen{Cardy:tv} and \citen{Lewellen:1991tb} for
details).
Among them, particularly important is the set referred to as 
Cardy's conditions\cite{Cardy:ir}, which 
require a certain
modular covariance of the annulus amplitudes determined from
boundary states in the set $\cB$.
These conditions are explained as follows.
Let us introduce arbitrary boundary states $\ket{B},\ket{B'}\in \cB$.
From 
these states,
we obtain a cylinder amplitude, 
\begin{equation}
\overlap{\bra{B}}{\ket{B'}}=\sum_{k}\fusion{k}{B'}{B}\chi_k(-1/\tau)\ ,  
\end{equation}
where $\fusion{k}{B'}{B}=\sum_{j}C_{B'}^j S_{jk}(C_B^j)^*$.
Here we have assumed that the characters $\chi_j$ are mapped 
to themselves under the 
modular S-transformation in the following manner:
\begin{equation}
\chi_j(\tau)=\sum_{k}S_{jk}\chi_k(-1/\tau)\ .  
\end{equation}
From generic properties expected in boundary CFTs, Cardy placed 
consistent conditions on the coefficients
$\fusion{k}{B'}{B}$.
These are called Cardy's conditions and are expressed as
\begin{align}
\fusion{k}{B'}{B}\in  \bbz_{\geq 0}\ ,\ \fusion{0}{B'}{B}=\delta_{BB'}\
 ,\ \forall \ket{B},\ket{B'}\in \cB\ ,
\end{align}
where $0$ in $\fusion{0}{B'}{B}$ denotes the
 Virasoro identity representation. 
Among these conditions, the particular ones
$\fusion{0}{B}{B}=1$ are necessary for the boundary states 
$\ket{B}$ to be fundamental.
In other words, boundary states cannot be fundamental if there are degeneracies of the Virasoro identity character in their self-overlaps.
At this point, we should mention whether Cardy's conditions are satisfied for
the set of boundary states constructed below. 
These boundary states will be explicitly expressed  
as linear 
combinations of the Virasoro Ishibashi states in the $\su2/G$ orbifold.
With the help of these expressions, it is not difficult to see that our set of boundary states
indeed does satisfy Cardy's conditions described above. (This is implicitly
explained in the Appendix.)

\section{Boundary states in $\su2/G$}\label{bssu2}

In this section, we construct bulk type and fractional type boundary
states in the $\su2/G$ orbifold.

\subsection*{Bulk type boundary states}

We begin with the
$\su2$ WZW model, in which there are a set of fundamental conformal boundary
states given (see, {\it e.g.}, Ref.~\citen{Callan:1994ub}) by
\begin{equation}
\ket{g}^{\su2}=2^{-1/4}\sum_{j\in \half\bbz_{\geq 0}}\sum_{m,n=-j}^j
D^j_{mn}(g^{-1})\dket{j;m,-n}^\Vir\ ,\ g\in SU(2)\ ,
\end{equation} 
where $D^j_{mn}(g)$ denotes the matrix
elements of the spin-$j$ representations of $SU(2)$, and
$\dket{j;m,n}^\Vir$ denotes
the Virasoro Ishibashi states 
of the conformal weights
$j^2$, which satisfy
\begin{equation}
\overlap{{}^\Vir\dbra{j;m,n}}{\dket{j';m',n'}^\Vir}=\delta_{j,j}\delta_{m,m'}\delta_{n,n'}\frac{1}{\eta(\tau)}(q^{j^2}-q^{(j+1)^2})\ ,
\end{equation}
where $\eta(\tau)$ denotes Dedekind's eta function.
These boundary states are mapped
as $\ket{g}^{\su2}\to \ket{g_L g g_R^{-1}}^{\su2}$
under $SU(2)_L\times SU(2)_R$, the symmetry of the
left-mover and right-mover modes of the $\su2$ WZW model.
This symmetry group includes an orbifolding group $G$ as a diagonal
subgroup.
Modding out the Hilbert space by the symmetry $G$, 
we obtain the untwisted sectors of the $\su2/G$ orbifold with
the charge conjugation modular invariant.\footnote{For details of 
rational orbifold
CFTs with the charge conjugate modular invariant, see Ref~\citen{Dijkgraaf:hb}.}
Thus, the $G$-invariant combinations,
\begin{equation}
 \ket{g}^G=\frac{1}{\sqrt{2|G|}}\sum_{h\in G}\ket{h g h^{-1}}^{\su2},\ g\in SU(2)\ \label{bulk}
\end{equation}
belong to these untwisted sectors.
Here note that the combinations $\ket{g}^G$ depend only 
on the $G$-orbits of $g$.
We can regard these combinations as fundamental conformal 
boundary states if the associated stabilizers, 
\begin{equation}
N(g)=\{h \in G\ | h g h^{-1}=g\}\ ,  
\end{equation} 
are
trivial, {\it i.e.}, $N(g)=\{1,-1\}$.
In this case, we call them bulk type boundary states.
However, if the stabilizers $N(g)$ are non-trivial, 
the expressions (\ref{bulk}) cannot represent  
fundamental boundary states.
This is because 
we find (by explicit calculation) $|N(g)|/2$-fold degeneracies of the Virasoro
identity character in their self-overlaps.
In this case, however, we can resolve these degeneracies by
decomposing the boundary states $\ket{g}^G$ into linear combinations of 
other fundamental boundary states.
Before demonstrating such a resolution, let us ask which $G$-orbits in $SU(2)$ have 
non-trivial stabilizers. 

\subsection*{Fixed orbits}

In the following, 
$G$-orbits in $SU(2)$ with non-trivial stabilizers are simply 
called fixed-orbits. 
We denote by $\cF$ the set of fixed orbits and denote 
by
$\overline{\cF}$ the supplement of $\cF$ in the set of all distinct
$G$-orbits in $SU(2)$. 
Prototypical fixed orbits are the conjugacy classes $C(G)$ 
(the $G$-orbits of the $G$-elements). 
These classes are introduced in the following manner.
First, we denote by $g_a(a=1,2,\cdots)$
and $g_b(b=1,2,\cdots)$ representatives of the distinct 
mutually non-conjugate $G$-elements and assume that 
the relations 
$g_a^{-1} \in [g_a],\ g_b^{-1}\notin [g_b]$ hold.
Here, we denote the $G$-orbits of $g\in SU(2)$ by 
$[g]=\{hgh^{-1}\ |\ h\in G\}$. 
We can always find the smallest positive
integers 
$n_a$ and $n_b$ satisfying $g_a^{n_a}=g_b^{n_b}=-1$. 
(Here, $a$ and $b$ are regarded as collections of
indices $1,2,\cdots$.)
Then, the $G$-elements $g_a$ and $g_b$ represent some classes in $C(G)$. 
All such classes are expressed as 
follows:
\begin{equation}
[\pm1];\ [g_a^s],\  s=1,\cdots,n_a-1;\ [\pm g_b^s],\  s=1,\cdots,n_b-1\ . 
\end{equation}
Here we note that the stabilizers of $[\pm
1],[g_a^s]$ and $[\pm g_b^s]$ are given, respectively, by 
\begin{align}
G,\ N_a=\{g_a^s\ |\ s=1,\cdots,2n_a \}\ \text{and}\ N_b=\{g_b^s\ |\
s=1,\cdots,2n_b \}. 
\end{align}
In addition to the above fixed orbits identified with $C(G)$, 
we can obtain the other fixed orbits
 by defining 
$2\times 2$ hermitian matrices $\sigma_a$ 
and $\sigma_b$ from the relations 
$g_a=e^{i\pi/n_a\sigma_a}$ and $g_b=e^{i\pi/n_b\sigma_b}$.
In this way, we can identify all the fixed orbits as in Table~\ref{fixed}.

\subsection*{The $\su2/G$ Ishibashi states}

We propose a resolution of $\ket{g}^G$ with $[g]\in \cF$, by expressing
$\ket{g}^G$ as linear combinations of other fundamental 
boundary states. 
Although it is difficult to construct the latter states
directly, we can guess
their expressions by studying other consistent boundary states, for
example, 
the Cardy states\cite{Cardy:ir}, in the $\su2/G$ orbifold.
In other words, we use the Cardy states to probe the space of generic
boundary states.
To obtain these Cardy states,
we need to identify 
the $\su2/G$ modular S-matrix and the $\su2/G$ Ishibashi states. 
The former was
obtained in Refs.~\citen{Cappelli:2002wq} and \citen{Dijkgraaf:hb}.
The latter are in one-to-one
correspondence with the chiral primary fields of $\su2/G$
\cite{Ishibashi:1988kg}.
We next review
 the $\su2/G$-primary-field 
content\cite{Dijkgraaf:hb,Cappelli:2002wq} before identifying the
$\su2/G$ Ishibashi states.

The distinct twisted sectors in the $\su2/G$ orbifold 
are in one-to-one
correspondence with $C(G)$. Here note that the untwisted sectors
correspond to the classes $[\pm 1]$.
Let us consider a class $[g]\in C(G)$ and the corresponding
$g$-twisted sector.  
The $g$-twisted sector is projected onto
irreducible representations of the stabilizer $N(g)$.
The set of these representations, denoted by $\cR(g)$, 
is chosen\cite{Dijkgraaf:hb}
as a suitable set of irreducible representations of $N(g)$.
To express $\cR(g)$, we need to explain some notation.
Let $\Irr(N(g))$ denote 
the set of all irreducible representations of $N(g)$.
Let us introduce two subsets of $\Irr(N(g))$ by 
\begin{equation}
\Irr^\pm(N(g))=\{r\in
\Irr(N(g))\ |\ r(-1)=\pm r(1)\}\ .
\end{equation}
Then, $\cR(g)$ is identified with 
either of $\Irr^\pm(N(g))$, as shown explicitly 
in Table~\ref{fixed}.
We assign 
the indices $(g,r)$ with $r\in \cR(g)$, to the
primary fields in the $g$-twisted sector.
In this way, all the chiral primary fields of $\su2/G$ are represented by
$(g,r)$, with $[g]\in C(G)$ and $r\in \cR(g)$.
As mentioned above, the primary fields $(g,r)$ correspond to
the $\su2/G$ Ishibashi states.
The corresponding states are 
denoted by $\dket{r}^{\su2/G}_{[g]}$ and are identified below. 

We can decompose each of the $\su2/G$ Ishibashi states 
into a linear combination of the Virasoro
Ishibashi states in the untwisted and twisted sectors. 
The Virasoro Ishibashi states in the untwisted
sectors were already introduced as $\dket{j;m,n}^\Vir$, which 
belong to the $1$-twisted and $-1$-twisted sectors, respectively, if
$j\in \bbz_{\geq 0}$ and $j\in \bbz_{\geq 0}+\half$. 
They are combined into the $\su2$ Ishibashi states of the spin-$0$ and
spin-$1/2$ highest-weight representations of the $\su2$ WZW model (the
parent theory of the orbifold) in the
following manner:
\begin{align}
 \dket{0}^{\su2}=\sum_{j\in\bbz_{\geq
 0}}\sum_{n=-j}^j\dket{j;n,-n}^\Vir,\   
\dket{1/2}^{\su2}=\sum_{j\in\bbz_{\geq 0}+\half}\sum_{n=-j}^j\dket{j;n,-n}^\Vir\ . 
\end{align}
The states $\dket{0}^{\su2}$ and $\dket{1/2}^{\su2}$ are projected onto 
the $\su2/G$ Ishibashi states of $(\pm 1,r)$ with $r\in \Irr^\pm(G)$, 
via suitable
projection operators\cite{Dijkgraaf:hb}. 
This projection can be expressed as follow:
\begin{equation}
 \dket{r}_{[\pm 1]}^{\su2/G}=\frac{\sqrt{\ch^{G}_r(1)}}{|G|}\sum_{h\in G}{\ch_r^G(h)}^*\sum_{j\in \half\bbz_{\geq 0}}\sum_{m,n=-j}^{j}D^j_{mn}(h)\dket{j;m,-n}^\Vir\ . \label{Ishi:untw}
\end{equation}
Here we have denoted by  $\ch^{G}_r$ the irreducible characters of $G$, with
$*$ representing the complex conjugate.
Next, we consider the Virasoro Ishibashi states in the twisted sectors to
decompose the $\su2/G$ Ishibashi states in these sectors.
The $\pm g_j^s$-twisted sectors with $j=a,b$ and
$s=1,\cdots,n_j-1$ are projected onto the representations $r\in \cR(\pm
g_j^s)$ of $(\pm g_j^s,r)$.
From the Virasoro characters of $(\pm g_j^s,r)$\cite{Dijkgraaf:hb,Cappelli:2002wq}, we can determine
which 
Virasoro Ishibashi states are contained in 
the $\su2/G$ Ishibashi states of $(\pm g_j^s,r)$.
In this way, we find the Virasoro Ishibashi
states in the twisted sectors as
\begin{equation}
\dket{m}_{[g_a^s]}^\Vir,\ m\in \bbz\ ;\ \ \dket{m}_{[g_b^s]}^\Vir,\ m\in
 \bbz\ ;\ \ \dket{m}_{[-g_b^s]}^\Vir,\ m\in \bbz+\half\ ,\label{TwVI} 
\end{equation}
where the states $\dket{m}_{[\pm g_j^s]}^\Vir$ are built up from the
(non-degenerate) Virasoro 
highest-weight states of conformal weights $(m+s/2n_j)^2$ in the $\pm g_j^s$-twisted
sectors and satisfy
\begin{equation}
 \overlap{{}^\Vir_{[\epsilon g_j^s]}\dbra{m}}{\dket{m'}_{[\epsilon' g_{j'}^{s'}]}^\Vir}
=\delta_{m m'}\delta_{[\epsilon g_j^s],[\epsilon' g_{j'}^{s'}]}\frac{1}{\eta(\tau)}q^{(m+s/2n_j)^2}\ ,
\end{equation}
where $\epsilon,\epsilon'=\pm 1$.
The Virasoro Ishibashi states $\dket{m}_{[\pm g_j^s]}^\Vir$ 
are combined into the $\su2/G$-Ishibashi
states in the $\pm g_j^s$-twisted sectors as follows:
\begin{align}
 \dket{r_a^+}_{[g_a^s]}^{\su2/G}&=\sum_{m\in \bbz}\dket{n_a
 m+r_a^+}_{[g_a^s]}^\Vir\label{Ishi:tw1}\ ,\\
 \dket{r_b^+}_{[g_b^s]}^{\su2/G}&=\sum_{m\in \bbz}\dket{n_b
 m+r_b^+}_{[g_a^s]}^\Vir\label{Ishi:tw2}\ ,\\
 \dket{r_b^-}_{[-g_b^s]}^{\su2/G}&=\sum_{m\in \bbz}\dket{n_b
 m+r_b^-+\half}_{[-g_a^s]}^\Vir\ .\label{Ishi:tw3}
\end{align} 
Here we have identified the representations $r_a^+\in \Irr^+(N_a)$ and 
$r_b^\pm \in
\Irr^\pm (N_b)$ with the positive integers $r_a^+=1,\cdots,n_a$ 
and
$r_b^\pm=1,\cdots,n_b$, so that the corresponding characters of $N_a$ and $N_b$ are expressed as
\begin{equation}
\ch^{N_a}_{r_a^+}(g_a)=e^{\frac{2\pi i}{n_a} r_a^+}\ ,\ 
\ch^{N_b}_{r_b^+}(g_b)=e^{\frac{2\pi i}{n_b}r_b^+}\ ,\ 
\ch^{N_b}_{r_b^-}(g_b)=e^{\frac{2\pi i}{n_b}(r_b^-+\half)}\ . 
\end{equation}
It is not difficult to see that the
expressions (\ref{Ishi:untw}) and (\ref{Ishi:tw1})--(\ref{Ishi:tw3})
reproduce the expected overlaps\cite{Ishibashi:1988kg},
\begin{equation}
 \overlap{{}^{\su2/G}_{\hspace{26pt}[g]}\dbra{r}}{\dket{r'}_{[g']}^{\su2/G}}=\delta_{(g,r),(g',r')}\chi_{(g,r)}(\tau)\ ,
\end{equation}
where $\chi_{(g,r)}$ denotes the Virasoro characters of the chiral primary fields
$(g,r)$, with $[g]\in C(G)$ and $r\in \cR(g)$.

\subsection*{Fractional type boundary states}

Defining the $\su2/G$-Ishibashi states as above 
leads to the unique set of the 
Cardy states (given later).
From these Cardy states, we can guess the expressions of more generic (fundamental) 
boundary states, as
mentioned in the Introduction.
Although such a prescription 
may not lead to a unique set of boundary states, we propose a
possible set below.
Each of the boundary states we propose corresponds to a
fixed orbit.
Conversely, an arbitrary fixed orbit $[g]\in \cF$ is associated with a
set of fundamental boundary states.  
These states are denoted by
$\ket{g;r}^G$ with $r\in \cR(g)$.
Here, the set $\cR(g)$ is identified with either $\Irr^+(N(g))$ or
$\Irr^-(N(g))$, as shown in Table~\ref{fixed}.
\footnote{Previously, we introduced the set $\cR(g)$ with $[g]\in C(G)$
to describe the primary-field content. Here we have not only
extended the definition of $\cR(g)$ for the case $[g]\in C(G)$ to the case
$[g]\in \cF$, but also changed the role of $\cR(g)$ to describe the boundary-state content.}
This identification results in the following relations for irreducible characters of
$N(g)$:
\begin{equation}
 \frac{|N(g)|}{2}=\sum_{r\in \cR(g)}(\ch_r^{N(g)}(1))^2\ . \label{cha:1}
\end{equation}
We
construct the states $\ket{g;r}^G$ to satisfy
\begin{equation}
 \ket{g}^G=\sum_{r\in \cR(g)}\ch_r^{N(g)}(1)\ket{g;r}^G\ . \label{reso_bs}
\end{equation}
Clearly, the right-hand side and the character relation (\ref{cha:1})
explain the origin of the $|N(g)|/2$-fold degeneracy associated with
the non-fundamental boundary state $\ket{g}^G$,
the left-hand side. 
\begin{table}[t]
 \caption{Several important
 quantities used in this note for a given $G$.  
}\label{fixed}
 \begin{equation*}
  \begin{array}{ccccc}
   &\text{Fixed orbits\ $\cF$} &\text{Conjugacy classes\ $C(G)$} &
    \text{Stabilizers}& 
     \\\hline\hline
  & [g] &[g]& N(g) & R(g) \\\hline
 &[\pm 1]&[\pm 1]& G & \Irr^\pm (G)\\
&
[e^{i\theta\sigma_a}],\ 0<\theta<\pi 
&[g_a^s],\ s=1,\cdots,n_a-1& N_a &
 \Irr^+(N_a)\\
&  [\pm e^{i\theta\sigma_b}],\ 0<\theta<\pi  
&[\pm g_b^s],\ s=1,\cdots,n_b-1
& N_b & \Irr^\pm(N_b)
  \end{array}
 \end{equation*}
\end{table}
The generic expressions of the states $\ket{g;r}^G$, 
with $[g]\in \cF$ and 
$r\in \cR(g)$, are given by
\begin{equation}
 \ket{g;r}^G=\frac{2\ch^{N(g)}_r(1)}{|N(g)|}\ket{g}^G+\ket{g;r}^G_\bt\ .
\label{ex1}
\end{equation}
Here, $\ket{g;r}^G_\bt$ denotes contributions to the states
$\ket{g;r}^G$ from the non-trivial 
twisted sectors in the $\su2/G$ orbifold.
To express these contributions, 
we define the following combinations of the Virasoro
Ishibashi states in the $\pm g_j^s$-twisted
sectors with $j=a,b$ and $s=1,\cdots,n_j$:
\begin{align}
 \ket{e^{i\theta\sigma_j}}^G_{[\pm
 g_j^s]}=\frac{2^{-1/4}}{\sqrt{n_j}}\sum_{m}e^{-i\theta(2m+s/n_j)}\dket{m}_{[\pm
 g_j^s]}^\Vir\ ,\ -\pi<\theta\leq\pi\ ,
\end{align}
where the indices $m$ are summed over the ranges given in (\ref{TwVI}).
For convenience, let us also introduce sets of
mutually non-conjugate $G$-elements:
\begin{align}
 \widetilde N_a=\{g_a^s\ |\ 1,\cdots,n_a-1\}\ , \   
\widetilde N_b=\{\pm g_b^s\ |\ 1,\cdots,n_b-1\}\ .\label{tilnb}
\end{align}
Using the quantities defined above, we can write 
the contributions $\ket{g;r}^G_\bt$, with $[g]\in \cF$ and $r\in \cR(g)$ 
(also see Table~\ref{fixed}),
as follows:
\begin{align}
& \hspace{7pt}\ket{\pm
 1;r}^G_\bt=\sum_a\sum_{h\in \widetilde
 N_a}{\ch^{G}_r(h)}^*\ket{1}^G_{[h]}+\sum_b\sum_{h\in \widetilde
 N_b}{\ch^{G}_r(h)}^*\ket{1}^G_{[h]}\ ,\label{contri1}\\
&\hspace{0pt}\ket{e^{i\theta\sigma_a};r}^G_\bt=
\sum_{h\in \widetilde N_a}
\left({\ch^{N_{a}}_r(h)}^*\ket{e^{i\theta\sigma_a}}^G_{[h]}+\ch^{N_a}_r(h)\ket{e^{-i\theta\sigma_a}}^G_{[h]}\right)\
 ,\label{contri2}\\
&\hspace{-9pt}\ket{\pm e^{i\theta\sigma_b};r}^G_\bt=\sum_{h\in\widetilde N_b}
{\ch^{N_{b}}_r(h)}^*\ket{e^{i\theta\sigma_b}}^G_{[h]}\ .\label{contri3}
\end{align}
Clearly, the above contributions are made only from non-trivial 
twisted-sector states.
This fact and the expressions (\ref{ex1}) imply that the boundary states
$\ket{g;r}^G$ have fractional masses, $2\ch^{N(g)}_r(1)/|N(g)|$, with
respect to the bulk type boundary states (\ref{bulk}).
Therefore, we can regard the states $\ket{g;r}^G$ as
fractional type orbifold boundary
states\cite{Oshikawa:1996dj,Diaconescu:1999dt}.

For clarity, we summarize the set of distinct boundary states 
constructed in this note (on the $\su2/G$ orbifold).
This set is given by 
\begin{equation}
 \cB=\{\ \ket{g}^G\ |\  [g]\in \overline{F}\ \}\cup \{\ \ket{g;r}^G\ |\ \ [g]\in F,\ r\in \cR(g)\ \}\ ,
\end{equation}
where the first (second) subset corresponds to the bulk type
(fractional type) boundary states. 
By construction, the Cardy states belong to the above set. 
The set of Cardy states is given by
\begin{equation}
 \{\ \ket{g;r}^G\ |\ [g]\in C(G),\ r\in\cR(g)\ \}\subset \cB\ ,
\, \label{Cardy_state}
\end{equation}
where the states $\ket{g;r}^G$ 
correspond to the Cardy states of the chiral 
primary
fields $(g,r)$.
This implies that the boundary states in the set (\ref{Cardy_state}) 
preserve the full $\su2/G$
symmetry\cite{Cardy:ir}. There are other classes of boundary states that preserve some
rational-symmetries in the $\su2/G$ orbifold. 
In Ref.~\citen{Cappelli:2002wq}, 
the boundary states that preserve
only the symmetry of the automorphism-type subalgebras of $\su2/G$ 
were constructed from
the Cardy states in the $\su2/\widetilde G$ orbifold, where
$\widetilde G$ denotes the groups extended from $G$ via the corresponding automorphism groups.  
Repeating the construction given there,
we can show that these symmetry-breaking states also belong to the set
$\cB$ and correspond to the conjugacy classes of $\widetilde G$.

\section*{Acknowledgements}
We would like to thank H. Ishikawa and M. Kato for helpful discussions.

\appendix

\section{Non-Vanishing Overlaps and Some Useful Identities}
Here we present several non-vanishing overlaps that 
contribute to the
cylinder amplitudes determined from the boundary
states given in this note. We also give useful identities for checking
Cardy's conditions in the set of these boundary states.
For these purposes, we first introduce some quantities.
Let us introduce generalized characters by
\begin{equation}
 \chi^{(n)}(\alpha|\tau)=\frac{1}{\eta(\tau)}\sum_{m\in\bbz}q^{(nm+\alpha)^2}\ .
\end{equation}
We denote the representations in $\Irr(N_a)$ and $\Irr(N_b)$ by the
positive integers $r_a=0,1,\cdots,2n_a-1$ 
and $r_b=0,1,\cdots,2n_a-1,$ so
that the corresponding characters have the following values at $g_a$ and $g_b$:
\begin{equation}
 \ch^{N_a}_{r_a}(g_a)=e^{i\pi  r_a/n_a}\ ,\ \ch^{N_b}_{r_b}(g_b)=e^{i\pi r_b/n_b}.
\end{equation}
Let us set $\ttau=-1/\tau$. Let us define a
function of $g\in SU(2)$ by 
\begin{equation}
\Delta(g)=\cos^{-1}\left(\half {\rm Tr}(g)\right)\ . 
\end{equation}
Then, basic overlaps are given as follows:
\begin{itemize}
 \item Overlap 1.
\begin{equation}
\overlap{{}^{\su2}\bra{g}}{\ket{g'}^{\su2}}=\chi^{(1)}(\Delta_1|\ttau)\ ,\  g,g'\in SU(2)\ , \ \Delta_1=\Delta(g {g'}^{-1})\ . 
\end{equation}
 \item Overlap 2. \\
Let us extend the left-hand side of
       (\ref{contri2}) to $0\leq\theta\leq \pi$ and $r_a\in \Irr(N_a)$ to define
       the corresponding combinations
       $\ket{e^{i\theta\sigma_a};r_a}^G_\bt$. Then, we have the overlaps,
\begin{align}
\overlap{{}^G_\bt\bra{e^{i\theta\sigma_a};r_a}}{\ket{e^{i\theta'\sigma_a};{r_a}'}_\bt^G}
&=\chi^{(n_a)}(\Delta_2^+|\ttau)+\chi^{(n_a)}(\Delta_2^-|\ttau)\nonumber\\
&\ \ \ -\frac{1}{n_a}\chi^{(1)}(\Delta_2^+|\ttau)-\frac{1}{n_a}\chi^{(1)}(\Delta_2^-|\ttau)\ ,
\end{align}
where $0\leq\theta,\theta'\leq\pi$, \ $r_a,{r_a}'\in \Irr(N_a)$ and $\Delta_2^\pm
=(\theta+\pi r_a)\pm (\theta'+\pi{r_a}')$.
 \item Overlap 3.\\
Similarly to the above case, we extend the definition of $\ket{\pm e^{i\theta\sigma_b};r_b}^G_\bt$ to cover the range $0\leq \theta\leq \pi$ and the representations $r_b\in \Irr(N_b)$. Then, we have the overlaps,
\begin{equation}
\overlap{{}^G_{\bt}\bra{e^{i\theta\sigma_b};r_b}}{\ket{e^{i\theta'\sigma_b};{r_b}'}^G_\bt}
=\chi^{(n_b)}(\Delta_3|\ttau)-\frac{1}{n_b}\chi^{(1)}(\Delta_3|\ttau)\ ,
\end{equation}
where $-\pi<\theta,\theta'\leq\pi$, \ $r_b,{r_b}'\in \Irr(N_b)$ and 
$\Delta_3=(\theta+\pi r_b)-(\theta'+\pi{r_b}')$.

\end{itemize}

By combining the above overlaps,  in principle, we can calculate all the cylinder
amplitudes obtained in our set of boundary states.
For such a calculation, however, 
it is useful to recognize the following facts (relations):
\begin{enumerate}
 \item Let us introduce $\ket{u}=\frac{2}{|N(g)|}\ket{g}^G$ and
       $\ket{v}=\frac{2}{|N(g')|}\ket{g'}^G$ for $g,g'\in SU(2)$. Then, it follows that
\begin{align}
\overlap{\bra{u}}{\ket{v}}&=\frac{2}{|N(g)|}\sum_{h\in
 [g']}\overlap{{}^{\su2}\bra{g}}{\ket{h}^{\su2}}\\
&=\frac{2}{|N(g')|}\sum_{h\in [g]}\overlap{{}^{\su2}\bra{h}}{\ket{g'}^{\su2}}\ .
\end{align}
 \item There exist $2\times 2$ matrices $\sigma_{am}$ with 
$m=1,2,\cdots,\ $ 
and $\sigma_{bm}$ with 
$m=1,2,\cdots$, such that the $G$-orbits $[\sigma_a]$
       and $[\sigma_b]$ are decomposed into 
$N_a$- and $N_b$-orbits as follows: 
\begin{align}
 [\sigma_a]&\to [\sigma_a]_{N_a}+
[-\sigma_a]_{N_a}+\sum_{m}[\sigma_{am}]_{N_a}\ ,\label{deco1}\\
 [\sigma_b]&\to
 [\sigma_b]_{N_b}+\sum_{m}[\sigma_{bm}]_{N_a}\ \label{deco2}.
\end{align}
Here note that the order of the orbits on the right-hand sides are given
       by 
       $|[\pm\sigma_a]_{N_a}|=|[\sigma_b]_{N_b}|=1$ and
       $|[\sigma_{am}]_{N_a}|=n_a,\ |[\sigma_{bm}]_{N_b}|=n_b$.
After some calculations, we can conclude that
\begin{align}
 \frac{1}{(n_a)^2}\overlap{{}^G\bra{e^{i\theta_a\sigma_a}}}{\ket{e^{i{\theta_a}'\sigma_a}}^G}&=\frac{1}{n_a}\chi^{(1)}(\Delta_a^+|\ttau)+\frac{1}{n_a}\chi^{(1)}(\Delta_a^-
 |\ttau)\nonumber\\&\hspace{75pt}+\sum_{m}\chi^{(1)}(\Delta_{am}|\ttau)\ ,\\
 \frac{1}{(n_b)^2}\overlap{{}^G\bra{e^{i\theta_b\sigma_b}}}{\ket{e^{i{\theta_b}'\sigma_b}}^G}&=
\frac{1}{n_b}\chi^{(1)}(\Delta_b|\ttau)+\sum_{m}\chi^{(1)}(\Delta_{bm}|\ttau)\
 ,
\end{align}
where 
\begin{align}
&&0\leq \theta_a,{\theta_a}'\leq\pi, &&& 
\Delta_a^\pm=\theta_a\pm {\theta_a}'&&\text{and}\   
\Delta_{am}=\Delta(e^{i\theta_a\sigma_a}e^{-i{\theta_a}'\sigma_{am}})\
 ;\\
&&-\pi<\theta_b,{\theta_b}'\leq \pi,&&& 
\Delta_b=\theta_b-{\theta_b}'&&\text{and}\  
 \Delta_{bm}=\Delta(e^{i\theta_b\sigma_b}e^{-i{\theta_b}'\sigma_{bm}})\ . 
\end{align}
 \item The contributions (\ref{contri1}) can be written 
as sums of the contributions (\ref{contri2}) and (\ref{contri3}), if 
the latter contributions are extended to $\theta=0,\pi$. 
This is because
       the irreducible characters of $G$  
are expressed in terms of irreducible characters of
       $N_a$ and $N_b$
for the $G$-elements in $\widetilde N_a$ and $\widetilde N_b$
       [defined in (\ref{tilnb})]. 
More concretely, for the representations $r\in\Irr(G)$,
there exist representations $r_a\in\Irr(N_a)$ and $r_b\in \Irr(N_b)$, and integers 
$c_r^a\ ({\rm or}\ {c_r^a}')$  and $c_r^b$, such that the following
       relations hold:
\begin{equation}
 \ch^G_r(h)=\begin{cases}
	     c_r^a(\ch_{r_a}^{N_a}(h)+{\ch_{r_a}^{N_a}(h)}^*),\ r_a\neq 0,n_a\ , &\\
\hspace{42pt}\text{or}\hspace{10pt} {c_r^a}'\ch_{r_a}^{N_a}(h),\  r_a=0,n_a& \text{for}\ h\in \widetilde N_a\ ,\\
	     c_r^b\ch^{N_b}_{r_b}(h)& \text{for}\ h\in \widetilde N_b\ ,\\
	    \end{cases}
\end{equation}
\begin{align}
 c_r^a&=\begin{cases}
	1 & \ch^G_r(1)=2 \mod n_a\ ,\\
	0 & \ch^G_r(1)=0 \mod n_a\ ,\\
	-1 & \ch^G_r(1)=-2 \mod n_a\ ,\\
       \end{cases}\\
 {c_r^a}'&=\begin{cases}
	1 & \ch^G_r(1)=1 \mod n_a\ ,\\
	0 & \ch^G_r(1)=0 \mod n_a\ ,\\
	-1 & \ch^G_r(1)=-1 \mod n_a\ ,\\
       \end{cases}
\\
c_r^b&=\begin{cases}
	1 & \ch^G_r(1)=1 \mod n_b\ ,\\
	0 & \ch^G_r(1)=0 \mod n_b\ ,\\
	-1 & \ch^G_r(1)=-1 \mod n_b\ .\\
       \end{cases}
\end{align} 
 \item The characters $\chi^{(n)}$ are related with the characters
       $\chi^{(1)}$ in the following way:
\begin{equation}
 \chi^{(1)}(\alpha|\tau)=\sum_{k=0}^{n-1}\chi^{(n)}(\alpha+k|\tau)\ .
\end{equation}
 \item Let us assign indices $\epsilon_r$ to the representations $r\in \Irr(G)$ as $\epsilon_r=\pm 1$
       for $r\in \Irr^\pm(G)$. Then, we can show the relations,
\begin{equation}
 \overlap{{}^G\bra{\epsilon_r\cdot 1;r}}{\ket{\epsilon_{r'}\cdot 1;r'}^G}=\sum_{r''\in \Irr(G)}\fusion{r'}{r''}{r}\chi_{(\epsilon_{r''}\cdot 1,r'')}(\ttau)\ ,
\end{equation}
where \begin{equation}
 \fusion{r'}{r''}{r}=\frac{1}{|G|}\sum_{h\in G}{\ch^G_r(h)}^*\ch^G_{r'}(h)
\ch^G_{r''}(h)\ . \label{fus}
\end{equation}
Note that the quantities (\ref{fus}) are the non-negative integers
       that appear in the decompositions of tensor-products of $r\in \Irr(G)$ and $r'\in \Irr(G)$:
\begin{equation}
(r)\otimes (r') =\sum_{r\in \Irr(G)}\fusion{r}{r'}{r''}(r'') \ .
\end{equation}
It may be useful to present $\chi_{(\epsilon_r\cdot 1,r)}$, 
the Virasoro characters of the $\su2/G$ primary fields 
$(\epsilon_r\cdot 1,r)$. These are explicitly expressed as follows:
\begin{equation}
 \displaystyle\chi_{(\epsilon_r\cdot 1,r)}(\tau)=\frac{1}{|G|}\frac{1}{\eta(\tau)}\sum_{h\in G}
{\ch^G_r(h)}^*\sum_{m\in\bbz}q^{m^2/4} e^{i m\Delta(h)}\ .
\end{equation}
\end{enumerate}


\begin{thebibliography}{99}
  


\bibitem{C_n}
C.~G.~Callan and I.~R.~Klebanov,
\PRL{72,1994,1968}; hep-th/9311092.

\bibitem{C_n2}
J.~Polchinski and L.~Thorlacius,
\PRD{50,1994,622}; hep-th/9404008.

\bibitem{Callan:1994ub}
C.~G.~Callan, I.~R.~Klebanov, A.~W.~W.~Ludwig and J.~M.~Maldacena,
\NPB{422,1994,417}; hep-th/9402113.

\bibitem{C_n3}
A.~Recknagel and V.~Schomerus,
\NPB{545,1999,233}; hep-th/9811237.

\bibitem{C_n4}
M.~R.~Gaberdiel, A.~Recknagel and G.~M.~Watts,
\NPB{626,2002,344}; hep-th/0108102.

\bibitem{C_n5}
M.~R.~Gaberdiel and A.~Recknagel,
\JHEP {11,2001,016}; hep-th/0108238.

\bibitem{C_n6}
R.~A.~Janik,
\NPB{618,2001,675}; hep-th/0109021.
\bibitem{C_n7}
L.~S.~Tseng,
\JHEP {04,2002,051}; hep-th/0201254.

\bibitem{Oshikawa:1996dj}
M.~Oshikawa and I.~Affleck,
\NPB{495,1997,533}; cond-mat/9612187.

\bibitem{Ginsparg:1987eb}
P.~Ginsparg,
\NPB{295,1988,153}.

\bibitem{Cappelli:2002wq}
A.~Cappelli and G.~D'Appollonio,
\JHEP {02,2002,039}; hep-th/0201173.

\bibitem{Ishikawa:2003xh}
H.~Ishikawa and A.~Yamaguchi,
\JHEP {04,2003,026}; hep-th/0301040.

\bibitem{Diaconescu:1999dt}
D.~E.~Diaconescu and J.~Gomis,
\JHEP{10,2000,001}; hep-th/9906242.

\bibitem{Cardy:ir}
J.~L.~Cardy,
\NPB{324,1989,581}.

\bibitem{aff}
I.~Affleck,
\JP{33,2000,6473}; cond-mat/0005286.

\bibitem{Ishibashi:1988kg}
N.~Ishibashi,
Mod.~Phys.~Lett.\ A~\andvol{4,1989,251}.

\bibitem{Cardy:tv}
J.~L.~Cardy and D.~C.~Lewellen,
\PLB{259,1991,274}.

\bibitem{Lewellen:1991tb}
D.~C.~Lewellen,
\NPB{372,1992,654}.

\bibitem{Dijkgraaf:hb}
R.~Dijkgraaf, C.~Vafa, E.~Verlinde and H.~Verlinde,
\CMP{123,1989,485}.


\end{thebibliography}
\end{document}